\documentclass[twocolumn,aps,prl,reprint,noshowpacs,letterpaper,floatfix,hypertext]{revtex4-2}
\usepackage{amsmath,amssymb}
\usepackage{txfonts}
\usepackage{bm}
\usepackage{textcomp}
\usepackage[pdftex]{graphicx,color}
\usepackage{braket}
\usepackage[]{hyperref}

\begin{document}
\title{Site Split of Antiferromagnetic $\alpha$-Mn Revealed by $^{55}$Mn Nuclear Magnetic Resonance}

\author{Masahiro Manago}
\email{manago@riko.shimane-u.ac.jp}
\author{Gaku Motoyama}
\affiliation{Department of Physics and Materials Science, Shimane University, Matsue 690-8504, Japan}
\author{Shijo Nishigori}
\affiliation{ICSR, Shimane University, Matsue 690-8504, Japan}
\author{Kenji Fujiwara}
\affiliation{Department of Physics and Materials Science, Shimane University, Matsue 690-8504, Japan}
\author{Katsuki Kinjo}
\author{Shunsaku Kitagawa}
\author{Kenji Ishida}
\affiliation{Department of Physics, Kyoto University, Kyoto 606-8502, Japan}
\author{Kazuto Akiba}
\author{Shingo Araki}
\author{Tatsuo C. Kobayashi}
\affiliation{Graduate School of Natural Science and Technology, Okayama University, Okayama 700-8530, Japan}
\author{Hisatomo Harima}
\affiliation{Department of Physics, Kobe University, Kobe 657-8501, Japan}

\begin{abstract}

    The magnetic structure of antiferromagnetic $\alpha$-Mn has been unclarified for almost 70 years
    since its magnetism was discovered.
    We measured the zero-field nuclear magnetic resonance spectra
    of antiferromagnetic $\alpha$-Mn to obtain further insight into magnetism below $T_{\text{N}} = 95$ K.
    The site II spectra split into two sites with five subpeaks owing to quadrupole interaction, and
    this shows that the ordered moments at site II are slightly tilted from the $[001]$ direction.
    The site III spectra revealed that this site splits into four sites below $T_{\text{N}}$.
    These findings clearly demonstrate that the antiferromagnetic $\alpha$-Mn symmetry is lower than previously considered.
\end{abstract}
\maketitle

Contrary to most elemental metals, $\alpha$-Mn, the stable form of Mn metal at ambient temperature and pressure,
has unique crystallographic properties.
It crystallizes into a complicated cubic structure without an inversion center ($I\bar{4}3m$, No.~217, $T_{d}^{3}$)
with four inequivalent Mn sites and 58 atoms in a unit cell.
$\alpha$-Mn exhibits antiferromagnetic (AFM) ordering at $T_{\text{N}} = 95$ K at ambient pressure.
Neutron diffraction measurements revealed the commensurate and non-collinear magnetic structure
in the AFM state.\cite{JPSJ.28.615,J.Appl.Phys.1.358024}.
The AFM phase is suppressed by the hydrostatic pressure at $\sim 1.5$ GPa,
and another magnetic phase emerges\cite{JMMM.310.e222,JPSJ.77.025001,JPSCP.30.011030,PhysRevResearch.2.043090}.
This is a weak ferromagnetic one with a small spontaneous magnetization and is suppressed at 4.2 GPa.
A remarkable feature is the anomalous Hall effect in the pressure-induced phase despite the absence of large spontaneous magnetization.
This effect is caused by the non-zero Berry curvature in the momentum space of non-collinear
antiferromagnets\cite{EuroPhysLett.108.67001}, which was observed in Mn$_3$Sn
and Mn$_3$Ge\cite{Nature.527.212,SciAdv.2.e1501870,PhysRevApplied.5.064009}.
A key factor of these rich properties in $\alpha$-Mn is the competition of AFM interactions
between neighbor Mn sites arising from its unique crystal structure.
Such competing interactions are also observed in magnetically-frustrated systems
$\beta$-Mn \cite{JPSJ.46.1754}, Y(Sc)Mn$_2$ \cite{JPSJ.62.1329}, and Mn$_3$P \cite{PhysRevLett.124.087202},
where application of hydrostatic pressure and atomic substitution can drastically alter the magnetic states.
The magnetism in $\alpha$-Mn has been studied for several decades, however, it still remains to be a fundamental subject.

The magnetic structure of $\alpha$-Mn remains elusive even at ambient pressure.
Neutron diffraction studies revealed that the magnetic moment is largest at site I,
followed by sites II, III, and IV\cite{JPSJ.28.596,J.Appl.Phys.1.358024}.
Sites III and IV split into two sites in the AFM state with a tetragonal distortion of the crystal structure\cite{JPSJ.28.615,J.Appl.Phys.1.358024}.
The site I magnetic moment points to $[001]$, and the site II moments slightly tilt from the $[001]$ axis,
showing the non-collinear orientation.
A previous nuclear magnetic resonance (NMR) study\cite{JPSJ.33.400}, conversely, revealed that the magnetic structure is more complicated than expected:
the NMR spectrum arising from site II split into two sites in the AFM state, and this suggested that site III split into four sites.
The recent NMR results \cite{JPSJ.90.085001,JPSJ.91.023709} agree with the previous report.
The site II split\cite{JPSJ.33.400}
suggests that the structure in the AFM state is lower than tetragonal\cite{PhysRevB.68.014407}.
However, it has not been taken into consideration in the analysis of the neutron study\cite{J.Appl.Phys.1.358024},
and thus, the true magnetic structure has not been revealed.
Because of the complex crystal structure, the magnetic structure of $\alpha$-Mn remains unclear for almost 70 years
since the antiferromagnetism of this system was first reported by neutron diffraction\cite{RevModPhys.25.100}.

In this paper, we report on the result of $^{55}$Mn zero-field NMR (ZF-NMR) measurements
on a high-quality $\alpha$-Mn sample at ambient pressure to obtain additional insight into the AFM structure.
It was identified that the two types of moments at split site II are parallel with a 6\% difference in size,
and the moments direction is slightly (6\textdegree) tilted from the $[001]$ axis,
in agreement with the previous neutron study.
We clarified that site III split into four sites in the AFM state,
with the hyperfine fields varying in 2--3 T.
In addition, site IV also splits into more than two sites with complicated spectra.
These findings confirm that the AFM $\alpha$-Mn symmetry is lower than that previously considered.
This presents a key indicator for AFM structure identification.

The $\alpha$-Mn sample was synthesized from the Pb-flux method in a horizontal configuration with a temperature gradient\cite{PhysRevMaterials.1.023402}
using Mn (99.999\%) and Pb (99.9999\%), as in Ref.~\onlinecite{PhysRevResearch.2.043090}.
A typical residual resistivity ratio between 2 and 300 K
is $\sim 17$ for samples by this method\cite{PhysRevResearch.2.043090},
which is higher than the previous values $\sim 2$\cite{JPSJ.77.025001,JMMM.310.e222}.
The sample was moderately crushed into powder in an argon atmosphere
with a typical grain size of $\sim 0.1$ mm.
The sample was inserted into the NMR coil with a typical diameter of 2 mm.
The standard spin-echo technique was used in the NMR measurements
without applying the external field.
Spontaneous moments in the AFM state induce hyperfine fields at the Mn sites, causing the Zeeman split
of the $^{55}$Mn nucleus.
The NMR measurements were performed at a temperature of 1.5 or 2 K well below $T_{\text{N}}$.

\begin{figure}
    \centering
    \includegraphics{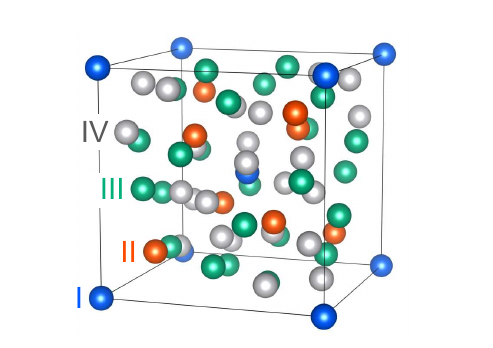}
    \caption{\label{fig:crystal}(Color online)
    Crystal structure of $\alpha$-Mn, drawn by VESTA\cite{JApplCryst.44.1272}.
    }
\end{figure}

\begin{table}
    \begin{center}
        \caption{\label{tab:sites}
            The local symmetry and parameters of quadrupole interaction of four Mn sites in
            $\alpha$-Mn for the paramagnetic state.
            Quadrupole parameters are calculated using band calculation.
        }
        \begin{tabular}{lllll}\hline\hline
            site & Wyckoff letter & symmetry    & $\nu_{\text{Q}}^{\text{calc}}$ (MHz) & $\eta^{\text{calc}}$ \\\hline
            I    & $2a$           & $\bar{4}3m$ & $0$                                  & -                    \\
            II   & $8c$           & $.3m$       & $2.520$                              & $0$                  \\
            III  & $24g$          & $..m$       & $2.627$                              & $0.882$              \\
            IV   & $24g$          & $..m$       & $1.808$                              & $0.851$              \\\hline\hline
        \end{tabular}
    \end{center}
\end{table}

Figure~\ref{fig:crystal} shows the crystal structure of $\alpha$-Mn.
Table \ref{tab:sites} summarizes the local symmetries of the Mn sites for the paramagnetic state
and calculated parameters of the electric field gradient (EFG) $V_{ij}$ for $^{55}$Mn by band calculation.
These values are used for the NMR spectra analysis described herein.
The EFG tensor is diagonalized by rotating the principal axes such that
$\lvert V_{zz} \rvert \ge \lvert V_{yy} \rvert \ge \lvert V_{xx} \rvert$,
and they are typically represented by the quadrupole frequency $\omega_{\text{Q}} \propto QV_{zz}$, and
the asymmetric parameter $\eta = \lvert V_{yy}-V_{xx} \rvert/\lvert V_{zz} \rvert$, where
$Q$ is the nuclear quadrupole moment.
The total Hamiltonian of the $^{55}$Mn nucleus ($I=5/2$) is expressed as follows:
\begin{equation}
    \mathcal{H} = -\gamma\hbar\bm{H}_{\text{hf}}\cdot\bm{I}
    +\frac{\hbar\omega_{\text{Q}}}{6} \left[ 3I_z^2 - I(I+1) + \frac{\eta}{2}(I_{+}^2 + I_{-}^2) \right],
\end{equation}
where $\gamma$ is the nuclear gyromagnetic ratio [$\gamma/(2\pi) = 10.554$ MHz/T], $\bm{H}_{\text{hf}}$
is the hyperfine field caused by the ordered moment.
Because site I in $\alpha$-Mn is located at the cubic symmetrical site, the EFG tensor is zero, leading to $\omega_{\text{Q}}=0$.
Under the hyperfine field, each Mn site, except for site I, splits into five lines owing to the finite $\omega_{\text{Q}}$.
The three-fold symmetry at site II ensures the asymmetric parameter is zero, and the $V_{zz}$ direction is along the
three-fold rotation axis.
Sites III and IV possess the same local symmetry, and
they were identified\cite{JPSJ.33.400}
based on the difference between ordered moments reported by neutron diffraction studies\cite{JPSJ.28.615,J.Appl.Phys.1.358024}.

\begin{figure}
    \centering
    \includegraphics{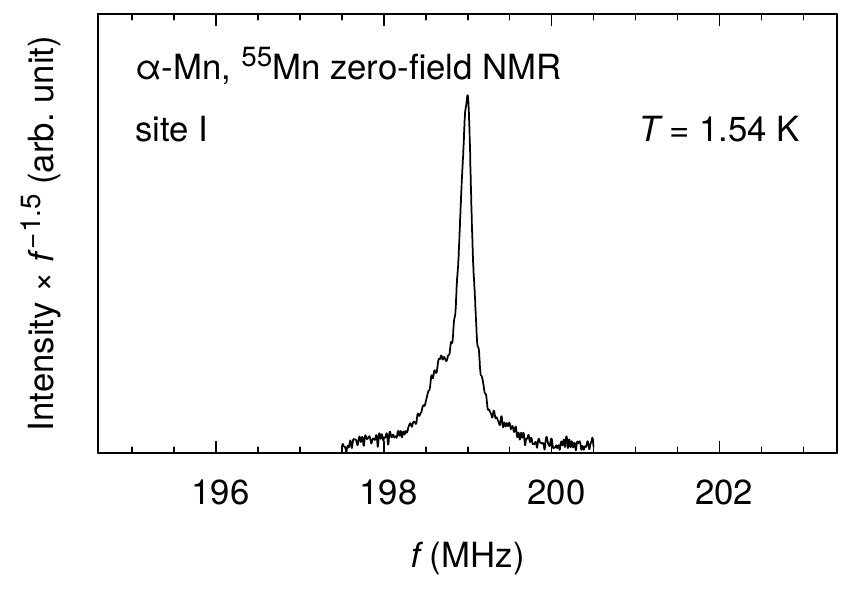}
    \caption{\label{fig:site-1}
        Zero-field NMR spectrum of $\alpha$-Mn originating from site I measured at 1.54 K in the AFM state.
        The intensity was multiplied by $f^{-1.5}$ to cancel the frequency ($f$) dependence of the emf and the skin effect (see text).
    }
\end{figure}

Figure \ref{fig:site-1} shows the ZF-NMR spectrum at site I.
The NMR intensity was multiplied by $f^{-1.5}$ to cancel the frequency ($f$) dependence of the emf
($f^2$) induced by nuclear moments and the skin effect ($f^{-0.5}$), which limits the effective sample volume.
This is essential for obtaining accurate intensity ratios of signals especially for sites III and IV shown later.
A sharp peak of 0.2 MHz width was observed at 199.0 MHz.
The absence of the quadrupole split is consistent with the local cubic symmetry of site I,
although the crystal structure should slightly distort below $T_{\text{N}}$ \cite{JPSJ.28.615,J.Appl.Phys.1.358024}.
The frequency is in agreement with previous reports\cite{JPSJ.33.400,JPSJ.90.085001}
and the width is approximately five times sharper than the previous report\cite{JPSJ.33.400},
showing good sample quality.
A minor broad peak was identified at 198.7 MHz, and this will be discussed later.

\begin{figure}
    \centering
    \includegraphics{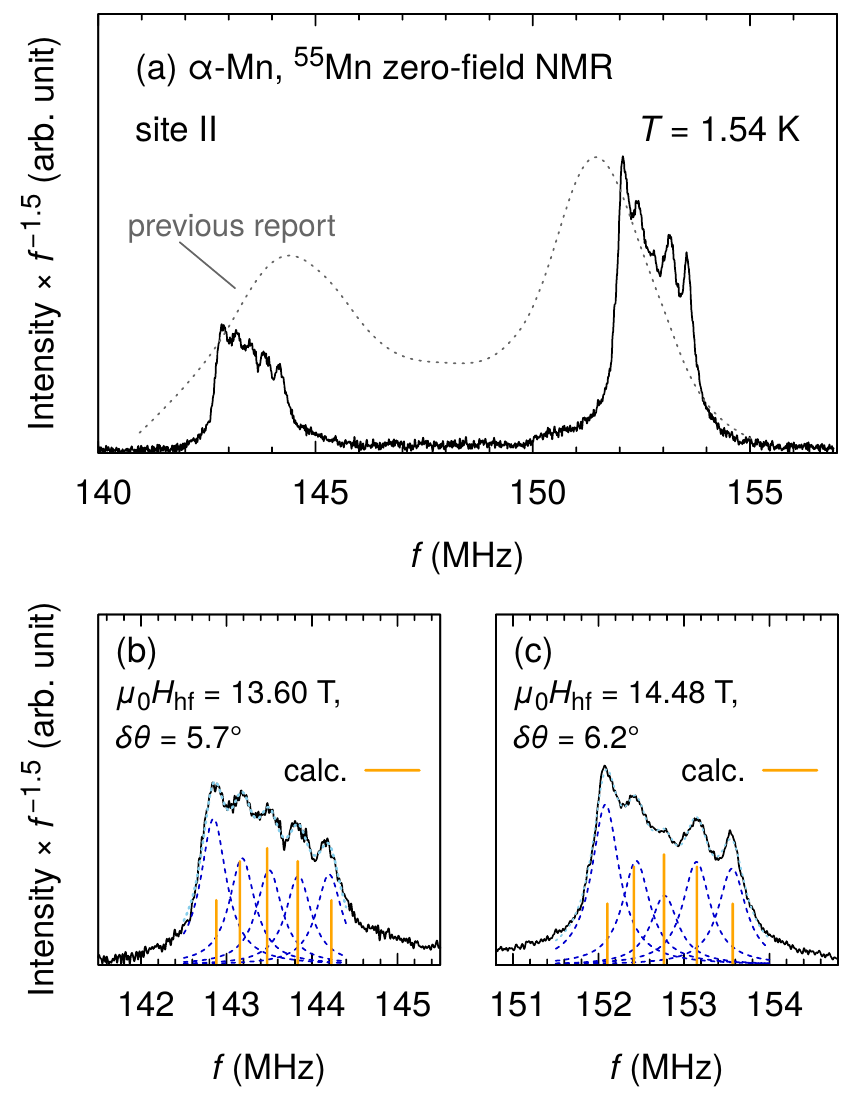}
    \caption{\label{fig:site-2}(Color online)
        (a) Zero-field NMR spectrum  of $\alpha$-Mn from site II measured at 1.54 K.
        The previous result by Yamagata and Asayama\cite{JPSJ.33.400} is shown with a dashed line for comparison.
        (b), (c) The analysis of the quadrupole split of site II for (b) the lower and (c) the higher lines.
        The experimental result was fitted with the summation of five Lorentz functions with the same width (0.4 MHz),
        as shown by the dashed lines.
        The hyperfine field $\mu_0 H_{\text{hf}}$ and its direction $\theta$ were obtained from the peak frequencies
        by numerical diagonalization, as shown by the vertical solid orange lines.
        $\delta\theta \equiv \theta-54.7$\textdegree\ indicates the angle of the hyperfine field from the $[001]$ direction.
    }
\end{figure}

Figure \ref{fig:site-2}(a) shows the ZF-NMR spectra at site II.
Two main peaks were observed in the ranges of 142--145 MHz and 151--154 MHz with five split lines
caused by the quadrupole interaction.
Because splitting is determined by the direction of hyperfine fields,
similar splitting in the two main peaks is the first direct evidence that moments at two Mn II sites are almost parallel.
The frequencies agree with previous reports\cite{JPSJ.33.400,JPSJ.90.085001},
and the two peaks are well separated in this result.

Figures \ref{fig:site-2}(b) and (c) show the site II spectra analysis.
The quadrupole split frequency depends on the angle $\theta$ between the magnetic field and
the EFG maximum principal axis $V_{zz}$.
Within the first-order perturbation with respect to the EFG, the frequency for
the $m-1 \leftrightarrow m$ transition is as follows:
\begin{equation}
    \nu_{m} = [\gamma/(2\pi) ] H_{\text{hf}}+(m-1/2)\frac{\nu_{\text{Q}}}{2} (3\cos^2 \theta-1),
\end{equation}
where $\nu_{\text{Q}} = \omega_{\text{Q}}/(2\pi)$.
The quadrupole splitting almost vanishes if $\theta = \cos^{-1} (1/\sqrt{3}) \simeq 54.7$\textdegree\ (so-called the magic angle).
This angle will be realized when the magnetic moment is along the $[001]$
because the $V_{zz}$ direction is along the $[111]$ at site II.
The spectral shape was reproduced with the summation of five Lorentz functions of 0.4 MHz width,
as shown by dashed lines in Figs.~\ref{fig:site-2}(b) and (c).
The best fit of peak frequencies was obtained with $\theta \simeq 61$\textdegree\ or $49$\textdegree,
which corresponds to the
angle $\delta\theta \equiv \lvert \theta-54.7\text{\textdegree} \rvert \sim 6$\textdegree\ from the $[001]$ directions.
This value is in excellent agreement with the previous neutron diffraction study by Lawson \textit{et al.} showing that the
moment at site II is in a non-collinear structure tilted by $6$\textdegree\
from the $[001]$ to $[110]$ directions\cite{J.Appl.Phys.1.358024}.
The angle somewhat differs from the report by Yamada \textit{et al.} \cite{JPSJ.28.615}.
Although it could not be distinguished from the NMR whether $\theta \simeq 61$\textdegree\ or $49$\textdegree,
i.e., whether the site II moment is tilted away from $[111]$ or toward $[111]$, the neutron results
indicate the latter.
The present NMR result shows that the directions of ordered moments at site II are similar
with only a difference of $\sim 6$\% in moment size.

\begin{figure}
    \centering
    \includegraphics{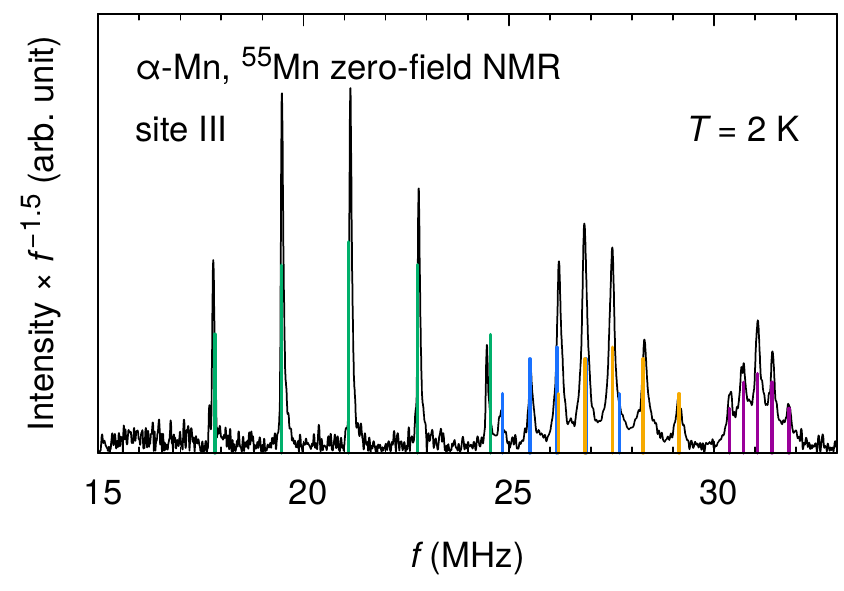}
    \caption{\label{fig:site-3}(Color online)
        Zero-field NMR spectrum of $\alpha$-Mn originating from site III measured at 2 K.
        The vertical lines show the calculated frequencies using the EFG parameters in Table~\ref{tab:sites}.
    }
\end{figure}

Figure \ref{fig:site-3} shows the ZF-NMR spectra at the site III region.
We identified 17 lines between 17--33 MHz, showing that the four inequivalent sites exist
in this frequency region with two sites almost overlapping between 25--30 MHz.
This result is the first direct experimental evidence that site III splits into four sites owing to magnetic ordering,
as suggested by the NMR spectra analysis \cite{JPSJ.33.400}.
The experimental result was reproduced using the EFG parameters shown in Table~\ref{tab:sites},
as shown by the vertical lines in Fig.~\ref{fig:site-3}.
The overlapped peaks between 25--30 MHz imply that the direction of
ordered moments are close for these two sites, as in site II, because the quadrupole splitting width
is determined by the angle between the hyperfine field and the $V_{zz}$ direction.
These sites may not be resolved by neutron diffraction.
By contrast, the other two sites at approximately 21 and 31 MHz are
quite different in internal field size.
Such a significant difference in some parts of site III is atypical and may represent key properties
for understanding its magnetism.

\begin{figure}
    \centering
    \includegraphics{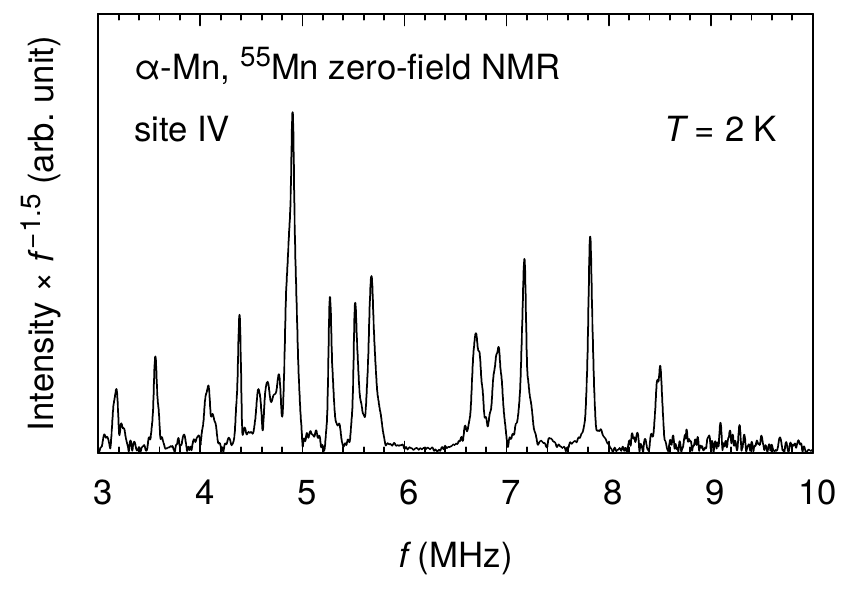}
    \caption{\label{fig:site-4}
        Zero-field NMR spectrum of $\alpha$-Mn of site IV at 2 K.
    }
\end{figure}

Figure \ref{fig:site-4} shows the ZF-NMR spectra of site IV in the 3--10 MHz range.
At least 16 lines were resolved in this range.
We consider that site IV also splits into four sites because of the identical local symmetry to site III in the paramagnetic
state.
Although detailed analysis of the site IV result is difficult,
this demonstrates that the site IV hyperfine field is approximately 0.5 T and significantly smaller
than that of site III.

The AFM $\alpha$-Mn symmetry has been considered as tetragonal $I_P\bar{4}2'm'$, which is the subgroup
of $I\bar{4}2m$ \cite{JPSJ.28.615,J.Appl.Phys.1.358024,Comment1}.
However, the present NMR results indicate that the symmetry is lower than this space group.
There is a possibility that orthorhombic distortion occurs, resulting in that the $\bar{4}$ local symmetry is broken
at site I, and the site II splits into two sites\cite{JPSJ.33.400,JPSJ.90.085001}.
More detailed symmetry consideration was recently reported by H. Fukazawa \textit{et al.}\cite{Fukazawa2022},
which is consistent with our scenario.
A problem is the signal intensity ratio of the split site II in the previous\cite{JPSJ.33.400,JPSJ.90.085001} and present studies:
the higher-frequency peak is stronger, as shown in Fig.~\ref{fig:site-2}(a), while the 1:1 split is expected for the orthorhombic symmetry.
Because the present NMR measurements cannot exactly identify the crystal structure,
precise X-ray or neutron diffraction measurements are required to determine the space group in the AFM $\alpha$-Mn.

The site II spectral shape is an atypical quadrupole-splitting signal
because the central peak is frequently the strongest among the split lines, however,
the satellite peaks are stronger for site II.
This spectrum discrepancy is probably caused by strong magnetic fluctuations.
The spin-echo intensity $M(2\tau)$ decays as the time interval $\tau$ between the first and second pulse oscillating fields
increases:
\begin{equation}\label{eq:exp}
    M(2\tau) = M(0)\exp(-2\tau/T_2).
\end{equation}
The values of the spin-echo decay rate $1/T_2$ differ in different lines
if the NMR line splits owing to quadrupole interaction.
In the simplest case, $1/T_2$ originating from magnetic fluctuations at the $m-1 \leftrightarrow m$ transition
will be as follows:\cite{PhysRevLett.19.146}
\begin{equation}
    \frac{1}{T_2} = \left[ I(I+1)-m(m-1) \right] \frac{1}{T_1},
\end{equation}
where $1/T_1$ is the nuclear spin-lattice relaxation rate, and its value
does not depend on where it is measured.
Thus, $1/T_2$ is fastest at the central peak ($-1/2 \leftrightarrow 1/2$ transition).
If the decay time $T_2$ is comparable to or shorter than $\tau$ because of strong fluctuations,
the intensity will be much weaker than that extrapolated to $\tau \to 0$.
This effect is the strongest at the central peak.

For site II, a rough estimation of $T_2$ was obtained from the $\tau$ dependence of the spectrum (not shown),
and $T_2 \sim 30$ \textmu s at the second satellite peaks at 1.54 K.
Because the spectrum in Fig.~\ref{fig:site-2} was measured by $2\tau = 26$ \textmu s,
the central peak may be weaker than the satellite peaks.
The double-horn shape identified in the recent NMR study\cite{JPSJ.90.085001} can be ascribed to this strong-fluctuation effect.
The value of $T_2 \sim 30$ \textmu s ($1/T_2 \sim 30$ ms$^{-1}$) is comparable to the nearly-AFM system $\beta$-Mn,
where $1/T_2$ exhibits anomalous temperature dependence $1/T_2 \sim T^{1/2}$ in the paramagnetic state
\cite{SolidStateCommun.16.1227,JPSJ.73.2305}.
The atypical spectral shape at site II in $\alpha$-Mn suggests that strong spin fluctuations
remain in the AFM state significantly below $T_{\text{N}}$.

The large $1/T_2$ was also observed at site I, and $1/T_2 \simeq 40$ ms$^{-1}$ was
obtained at 1.6 K from the fitting with Eq.~(\ref{eq:exp}).
The magnetic moment at site I strongly fluctuates even significantly below $T_{\text{N}}$.
Conversely, $1/T_2$ is smaller at the minor peak measured at 198.75 MHz, and $1/T_2 \simeq 20 $ ms$^{-1}$.
Although the minor peak origin remains unclear, the relative
intensity of this peak is $\sim 0.1$ times of the main peak for $\tau \to 0$.

\begin{table}
    \begin{center}
        \caption{\label{tab:fields}
        The hyperfine fields at the $^{55}$Mn sites at 2 or 1.5 K in the AFM state.
        The values for site IV are tentative; they strongly depend on the assignments of the spectra and the values
        of the quadrupole parameters $\nu_{\text{Q}}$ and $\eta$ used for the analysis.
        }
        \begin{tabular}{lllll}\hline\hline
            site & \multicolumn{4}{c}{hyperfine fields (T)}                         \\\hline
            I    & 18.86                                                            \\
            II   & 13.68                                    & 14.48                 \\
            III  & 1.995                                    & 2.476 & 2.607 & 2.938 \\
            IV   & 0.43                                     & 0.47  & 0.47  & 0.69  \\\hline\hline
        \end{tabular}
    \end{center}
\end{table}

\begin{figure}
    \centering
    \includegraphics{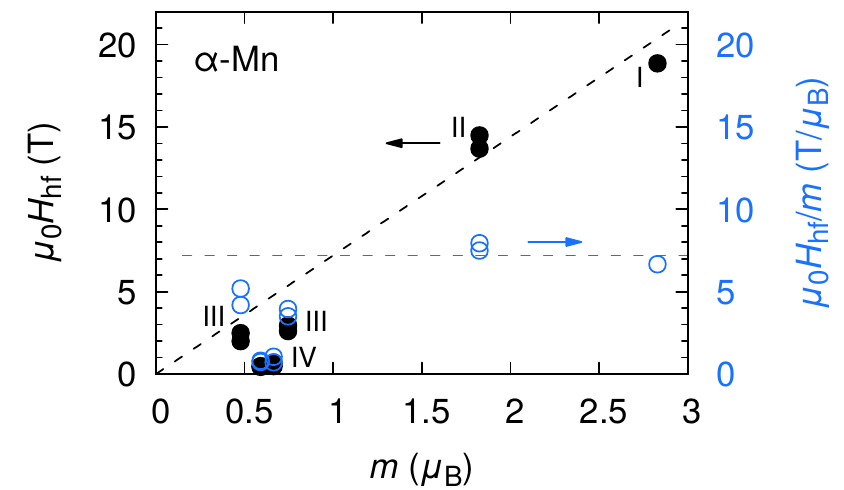}
    \caption{\label{fig:hyperfine}(Color online)
    The hyperfine field against the ordered moment of $\alpha$-Mn (left axis) and the
    field divided by the moment, or apparent hyperfine coupling constant (right).
    The values of the ordered moments are from Ref.~\onlinecite{J.Appl.Phys.1.358024}.
    }
\end{figure}

The intensities of the hyperfine fields for the Mn sites are summarized in Table~\ref{tab:fields}.
The relationship between the hyperfine fields and ordered moments reported by Lawson \textit{et al.}\cite{J.Appl.Phys.1.358024}
is shown in Fig.~\ref{fig:hyperfine}.
As asserted in the previous paper\cite{JPSJ.33.400}, the hyperfine field is not proportional
to the spontaneous moments, which is significant at sites III and IV.
Three possible mechanisms may cause this nonlinear relationship\cite{JPSJ.33.400}.
The first is related to the analysis of the neutron diffraction.
The form factors for the Mn sites used in the analysis are common to each other\cite{JPSJ.28.615,J.Appl.Phys.1.358024}.
If the actual form factors vary in different sites, magnetic moments may differ from those obtained by the analysis.
It should be also noted that the sites II and III split in the present NMR spectra are not considered in the analysis.
The second is the site dependence of the hyperfine coupling constant.
The contribution of the $4s$ electron to the ordered moments, or the $s$-$d$ mixing, may vary in different sites.
The hyperfine coupling constant is negative and large for the $d$ electron owing to core polarization,
whereas the $s$ electron provides a positive coupling constant from the contact interaction.
In metallic vanadium,
cancellation of the hyperfine field occurs between $s$ and $d$ electrons,
leading to a small $^{51}$V NMR Knight shift reduction in the superconducting state despite the $s$-wave symmetry\cite{RevModPhys.36.177,JETPLett.87.316}
However, this effect would not be a major contribution for $^{55}$Mn because $d$ electrons play a significant role in $\alpha$-Mn.
The third is the hyperfine field from neighbor Mn sites.
This effect will be relatively larger as on-site moments are smaller.

In summary, ZF-NMR measurements were performed on $\alpha$-Mn in the AFM state, and it was revealed that
site II splits into two sites with slightly tilted ($\sim 6$\textdegree\ from $[001]$ direction) magnetic moments.
The two split moments are almost parallel with similar moment size.
The NMR spectra also revealed that site III splits into four sites.
More than two sites were identified in the site IV region.
These findings demonstrate that the AFM $\alpha$-Mn symmetries are lower than previously proposed by
neutron measurements.

\begin{acknowledgments}
    This paper is dedicated to Prof.~K. Asayama, who made a significant contribution
    to the development of condensed matter physics based on NMR and nuclear quadrupole resonance measurements.
    The authors would like to extend their gratitude to H. Fukazawa and Y. Kohori for their insightful discussions.
    One of the authors (K. F.) thanks K. Asayama for the stimulating encouragements.
    K. F. and M. M. thank K. Miyoshi, M. Ohno, C. Akasaka, and W. Irie for their experimental supports.
    This work was supported by the JSPS KAKENHI Grant Numbers JP21H01042, JP20H00130, JP21K03448,
    and FY2021 Shimane University Internal Competitive Grants.
    One of the authors (K. K.) was also supported by and the JST SPRING (Grant Number JPMJSP2110).
\end{acknowledgments}


\begin{thebibliography}{28}
\bibitem [{Com()}]{Comment1}%
  \BibitemOpen
  \href@noop {} {}\bibinfo {note} {There is an erratum in the symbol for the
  magnetic space group in the paper by Lawson \textit{et al.} $I_P\bar{4}2'm'
  \equiv P_I\bar{4}2_1c $ is the correct one for the spece group proposed by
  them.}\BibitemShut {Stop}%
\end{thebibliography}
\end{document}